\documentclass[pra,twocolumn, preprintnumbers,amsmath,amssymb, superscriptaddress, nofootinbib]{revtex4}

\usepackage{graphicx, color, graphpap}
\usepackage{amsmath}
\usepackage{enumitem}
\usepackage{amssymb}
\usepackage{amsthm}
\usepackage{pstricks}
\usepackage{float}
\usepackage{multirow}
\usepackage{hyperref}
\usepackage{overpic}
\usepackage[T1]{fontenc}
\usepackage{algorithm}
\usepackage{algpseudocode}
\input{Qcircuit}


\long\def\ca#1\cb{} 

\newcommand{\CZ}{\mathrm{CZ}}

\newcommand{\ketbra}[2]{| \hspace{1pt} #1 \rangle \langle #2 \hspace{1pt} |}

\newcommand{\ip}[2]{\langle #1|#2\rangle}      

\newcommand{\AC}{\mathcal{A}}

\newcommand{\Tr}{{\rm Tr}}

\renewcommand{\geq}{\geqslant}

\newcommand{\ad}{^\dagger}

\newtheoremstyle{example}{\topsep}{\topsep}%
{}
{}
{\bfseries}
{:}
{   }
{\thmname{#1}\thmnumber{ #2}}
\theoremstyle{example}

\theoremstyle{definition}

\newcommand\numberthis{\addtocounter{equation}{1}\tag{\theequation}}

\begin{document}

\title{Learning the quantum algorithm for state overlap}

\author{Lukasz Cincio} 
\affiliation{Theoretical Division, Los Alamos National Laboratory, Los Alamos, NM 87545, USA.}

\author{Yi\u{g}it Suba\c{s}\i} 
\affiliation{Theoretical Division, Los Alamos National Laboratory, Los Alamos, NM 87545, USA.}

\author{Andrew T. Sornborger} 
\affiliation{Information Sciences, Los Alamos National Laboratory, Los Alamos, NM 87545, USA.}

\author{Patrick J. Coles} 
\affiliation{Theoretical Division, Los Alamos National Laboratory, Los Alamos, NM 87545, USA.}

\begin{abstract}
Short-depth algorithms are crucial for reducing computational error on near-term quantum computers, for which decoherence and gate infidelity remain important issues. Here we present a machine-learning approach for discovering such algorithms. We apply our method to a ubiquitous primitive: computing the overlap $\Tr(\rho\sigma)$ between two quantum states $\rho$ and $\sigma$. The standard algorithm for this task, known as the Swap Test, is used in many applications such as quantum support vector machines, and, when specialized to $\rho = \sigma$, quantifies the Renyi entanglement. Here, we find algorithms that have shorter depths than the Swap Test, including one that has a constant depth (independent of problem size). Furthermore, we apply our approach to the hardware-specific connectivity and gate sets used by Rigetti's and IBM's quantum computers and demonstrate that the shorter algorithms that we derive significantly reduce the error - compared to the Swap Test - on these computers.
\end{abstract}

\break
\newpage
\newpage
\maketitle

\section{Introduction}\label{sctintro}

Quantum supremacy \cite{preskill2012quantum} may be coming soon \cite{neill2017blueprint}. While it is an exciting time for quantum computing, decoherence and gate fidelity continue to be important issues \cite{ball2018the}.  Ultimately these issues limit the depth of algorithms that can be implemented on near-term quantum computers (NTQCs) and increase the computational error for short-depth algorithms. Furthermore, NTQCs do not currently have enough qubits and sufficient gate fidelities to fully leverage the benefit of quantum error-correcting codes \cite{fowler2012surface, you2013simulating}. This highlights the need for general methods to reduce the depth of quantum algorithms in order to avoid the accumulation of errors.

Analytical efforts to find short-depth algorithms face several challenges. First, quantum algorithms are fairly non-intuitive to classically trained minds. Second, actual NTQCs may not be fully connected. Third, different NTQCs use different fundamental gate sets. It may not be obvious how to optimize algorithms for a given connectivity and a given gate set. This motivates the idea of an automated approach for discovering and optimizing quantum algorithms \cite{benedetti2018generative, mitarai2018quantum, khaneja2005optimal, gepp2009review, lukac2003evolutionary, venturelli2018compiling, khatri2018quantum, haner2018software, maslov2017basic, martinez2016compiling, chong2017programming, swaddle2017generating, zahedinejad2016designing, las2016genetic}.

An analogous problem in classical computing, known as logic synthesis, has a relatively longer history and has been extensively studied \cite{hachtel2006logic}. Machine-learning methods have been used in this context. For instance Ref.~\cite{haaswijk2018deep} shows how logic optimization algorithms can be discovered automatically through the use of deep learning.

\begin{figure}
\begin{center}
 \includegraphics[width=\columnwidth]{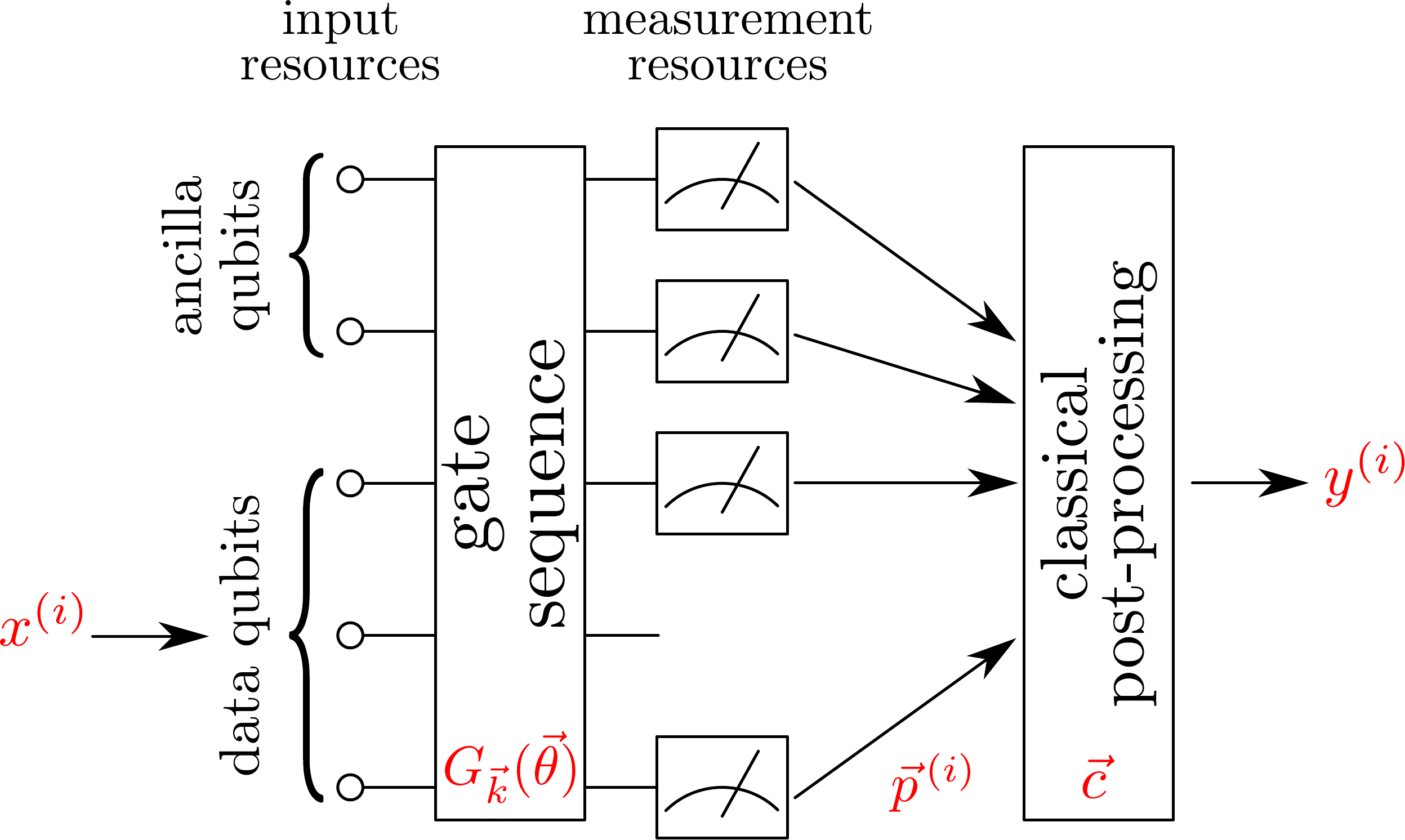}
\caption{Machine-learning approach to discovering and optimizing quantum algorithms. We optimize an algorithm for a given set of resources, which includes input resources (ancilla and data qubits) and measurement resources (i.e., which qubits can be measured). The algorithm is then determined by the quantum gate sequence and the classical post-processing of the measurement results. To find the algorithm that computes the function $x \to f(x)$, we minimize a cost function that quantifies the discrepancy between the desired output $f(x^{(i)})$ and the actual output $y^{(i)}$ for a set of training data inputs $\{x^{(i)} \}$. If the training data are sufficiently general, the algorithm that minimizes the cost should be a general algorithm that computes $f(x)$ for any input $x$.}
\label{figure1}
\end{center}
\end{figure}

In this work, we take a machine-learning approach to developing quantum algorithms, see Fig.~\ref{figure1}. Our approach can be applied either to ideal hardware to derive fundamental algorithms or to a non-fully connected hardware with a non-ideal gate set to derive hardware-specific algorithms. We conceptually divide a quantum computation into the available resources, consisting of input qubits (data qubits and ancilla qubits) and output measurements, and the algorithm, consisting of a quantum gate sequence and classical post-processing of the measurement results (see Fig.~\ref{figure1}). Fixing the resources as hyperparameters, we optimize the algorithm in a task-oriented manner, i.e., by minimizing a cost function that quantifies the discrepancy between the algorithm's output and the desired output. The task is defined by a training data set that exemplifies the desired computation. Our machine-learning approach is used to discover small algorithm instances that can be later manually generalized to arbitrary problem size.

We emphasize that our work goes beyond quantum compiling, which has received recent attention \cite{venturelli2018compiling, haner2018software, khatri2018quantum, maslov2017basic, martinez2016compiling, chong2017programming}. 
Quantum compiling corresponds to finding a hardware-specific gate sequence that generates the same unitary as a high-level gate sequence defined for an idealized hardware. 
Various techniques have been employed in these works such as temporal planning (e.g. \cite{venturelli2018compiling}).
Machine-learning techniques have also been used to decompose small scale unitaries into one and two-body gates ~\cite{swaddle2017generating,zahedinejad2016designing}. Although our method can be used in this way to optimally compile a known unitary or gate sequence, our main goal is to discover novel algorithms via task-oriented programming. 

Other automated algorithm-discovery approaches have been employed in the literature. Gepp and Stocks \cite{gepp2009review} review much of the early work to evolve quantum algorithms using genetic programming such as \cite{lukac2003evolutionary} (for more recent work see for example \cite{las2016genetic}). In these approaches the gate set is typically discrete. An alternative approach is to define an ansatz or template for the quantum circuit composed of gates that depend on continuous parameters. The circuit is then trained to perform a given task by tuning these parameters \cite{mitarai2018quantum, benedetti2018generative}. Our approach is distinct from previous works in that we do not start with an ansatz or template for the quantum circuit; nor do we restrict to a discrete gate set as is usually done in algorithms based on genetic programming. In this sense our approach combines desirable aspects of the two types of approaches in the literature.

We apply our approach to a ubiquitous task: computing the overlap between two quantum states. This computation yields $|\ip{\psi}{\phi}|^2$ for two pure states $\ket{\psi}$ and $\ket{\phi}$, and more generally gives $\Tr(\rho\sigma)$ for two (possibly mixed) states $\rho$ and $\sigma$. Furthermore, when specialized to the case $\rho = \sigma$, it computes the purity $\Tr(\rho^2)$ of a given state $\rho$.

\begin{figure}
\begin{center}
 \includegraphics[width=\columnwidth]{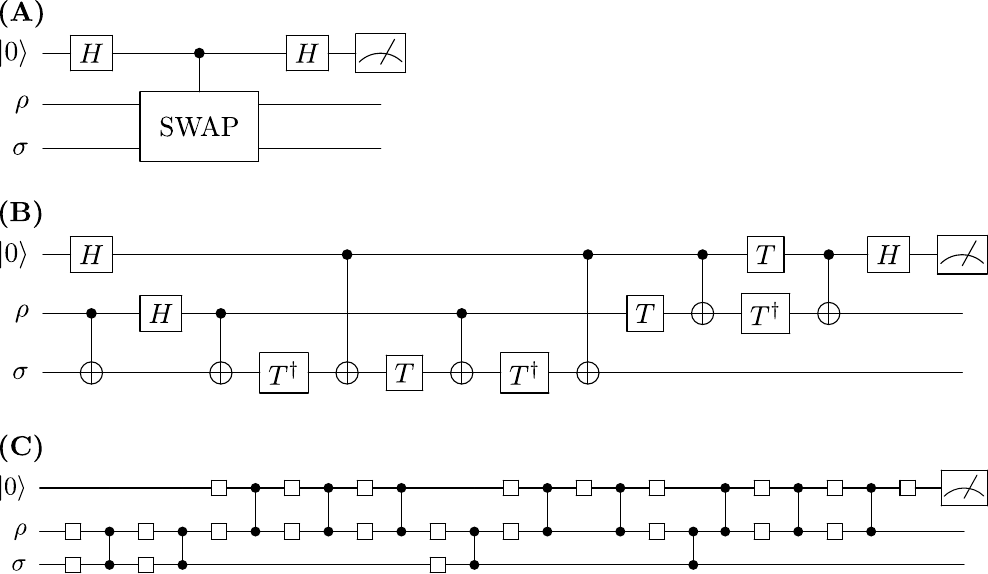}
\caption{Swap Test circuits. (A) The canonical Swap Test circuit. $H$ indicates the Hadamard gate. (B) The Swap Test circuit adapted for IBM's 5-qubit quantum computer, constructed by decomposing controlled-swap into the Toffoli gate, via Refs.~\cite{smolin1996five, shende2009cnot}, and then manually eliminating gates that had no effect on the output. $T$ is the $\pi/8$ phase gate. (C) The structure of a Swap Test circuit, showing the locations of the one-qubit gates and controlled-$Z$ gates, constructed automatically by Rigetti's compiler for their 19-qubit quantum computer. Appendix \ref{sctdetails} gives the full specification of that circuit.}
\label{figure2}
\end{center}
\end{figure}

There is a well-known algorithm for this task called the Swap Test \cite{buhrman2001quantum, gottesman2001quantum}. In quantum optics the Swap Test has a simple physical implementation \cite{garcia2013swap, patel2016quantum, ferreyrol2013implementation}. However, for gate-based quantum computers (e.g., IBM's, Google's, and Rigetti's superconducting quantum computers and IonQ's trapped-ion quantum computer), the optimal implementation of the Swap Test is not obvious, and for single qubit states involves 14 and 34 gates for IBM's 5-qubit and Rigetti's 19-qubit quantum computer respectively, see Fig.~\ref{figure2}. Larger gate count for Rigetti's computer is mainly due to its lower connectivity.  For example, the Swap Test was experimentally implemented on a 5-qubit computer based on trapped ions \cite{linke2017measuring} to quantify entanglement, with an algorithm employing $7$ two-qubit gates and $11$ one-qubit gates. Figure~\ref{figure2}(B) and (C) respectively show decompositions of the Swap Test for IBM's and Rigetti's quantum computers \cite{smith16quil, cross17ibm}. This highlights the non-trivial nature of implementing the Swap Test algorithm.

Here, our machine-learning approach finds algorithms with a shorter depth than the Swap Test for computing the overlap. We do this by initially specializing the training data to one- and two-qubit states and then manually generalizing the resulting algorithms to input states of arbitrary size. We first consider the same ``quantum resources'' as the Swap Test (access to a qubit ancilla and measurement on the ancilla), and our approach reduces the gate count to $4$ controlled-NOTs ($\text{CNOT}$s) and $4$ one-qubit gates. We call this our Ancilla-Based Algorithm (ABA). Then we allow for the additional resource of measuring all of the qubits, which gives an even shorter depth algorithm that essentially corresponds to a Bell-basis measurement with classical post-processing. We call this our Bell-Basis Algorithm (BBA). This algorithm has a constant depth of two gates, while the classical post-processing scales linearly in the number of qubits of the input states. In that regard, our machine-learning approach independently discovered the algorithm of Garcia-Escartin and Chamorro-Posada for computing state overlap \cite{garcia2013swap}. We also find short-depth algorithms for the specific hardware connectivity and gate sets used by IBM's and Rigetti's quantum computers, which is crucial for reducing the computational error. Indeed, we found that our short-depth algorithms reduced the root-mean-square error (compared to the Swap Test) by $66 \%$ on IBM's 5-qubit computer and by $70 \%$ on Rigetti's 19-qubit computer.

Due to the fundamental nature of computing state overlap, the Swap Test appears in many applications. In quantum supervised learning \cite{lloyd2013quantum, wiebe2014quantum}, which subsumes quantum support vector machines \cite{rebentrost2014quantum}, the Swap Test is used to assign each data vector to a cluster. The Swap Test allows one to quantify entanglement for many-body quantum states \cite{linke2017measuring, johri2017entanglement} using the Renyi order-2 entanglement, given by $H^{(2)} = - \log \Tr (\rho^2)$.
The Swap Test is useful for benchmarking on a quantum computer, since it can quantify the purity $\Tr(\rho^2)$ and hence the amount of decoherence that has occurred. For all of the above applications, one of our shorter-depth algorithms can be directly substituted in place of the Swap Test.

Note that if $\rho$ and $\sigma$ represent states on $n$ qubits, the difficulty for computing $\Tr(\rho\sigma)$ scales exponentially with $n$ for a classical computer. In contrast, the Swap Test has a circuit depth that grows linearly in $n$, giving an exponential speedup. Our ABA also has this property of scaling linearly with $n$, and it reduces the number of gates in the circuit by a factor of $\sim 2.3$ (relative to the Swap Test circuit decomposed in terms of CNOTs, as shown in Fig. \ref{figure2}(B)). On the other hand, our BBA has the nice feature that its circuit depth is constant, independent of $n$ (although the complexity of its classical post-processing grows linearly in $n$). Due to its constant circuit depth, the BBA seems to be the best algorithm for quantifying state overlap on NTQCs.

In what follows, we first present our machine-learning approach for discovering quantum algorithms. This approach can be used to find other algorithms besides the one that computes the overlap and hence should be of independent interest. We also give the full details of the approach and discuss its scaling with various resources. Next, we present our main results: short-depth circuits for computing state overlap on idealized hardware. Then, we present hardware-specific algorithms for computing overlap. Finally we discuss our implementation of these algorithms on Rigetti's and IBM's quantum computers, leading to a reduction in the computational error relative to the Swap Test.

\section{Machine-Learning Approach}\label{sctapproach}

Our machine-learning approach is summarized in Fig.~\ref{figure1}. The variables are divided up into the hyperparameters (i.e. the ``resources'') and the optimization parameters (i.e. the ``algorithm'').

\subsection{Resources}

The hyperparameters are the quantum resources of the circuit. At the input, the resources are the number of ancilla qubits and data qubits that store the input data for the computation. At the output, the resources are the locations of the measurements (see Fig.~\ref{figure1}). As an example, in the Swap Test for single-qubit states, we are allowed access to one ancilla qubit and two data qubits at the input, and we can measure only the ancilla qubit at the output. 

The input data may be classical or quantum, depending on the computation of interest. In the case of state overlap, the input data are quantum states and hence no encoding is necessary. However, for completeness, we note that our approach also applies to classical inputs, in which case the encoding (i.e., storing the classical data in the quantum state of the data qubits) can be treated as a hyperparameter that one fixes while optimizing the algorithm.

\subsection{Algorithm}

Our approach searches for an optimal algorithm, where we consider the algorithm to be a quantum gate sequence with associated classical post-processing. We parameterize (and hence optimize over) both the gate sequence and the post-processing.

Let us first consider the gate sequence. We define a gate set $\AC = \{ A_j (\theta)\}$. Here, each gate $A_j$ is either a one-qubit or two-qubit gate and may also have an internal continuous parameter $\theta$. Hence, $\AC$ is a discrete set, but each element of $\AC$ may have a continuous parameter associated with it. The precise choice of $\AC$ depends on which hardware one is considering. For example, the connectivity differs between IBM and Rigetti hardware, and the former employs $\text{CNOT}$ gates while the latter employs controlled-$Z$ gates. For IBM's 5-qubit computer ``ibmqx4'' we can write out the gate set as
\begin{align}
\label{eqn1}
\AC_{\text{ibmqx4}} = \{ & \text{CNOT}^{10}, \text{CNOT}^{20}, \text{CNOT}^{21}, \text{CNOT}^{32}, \notag\\
& \text{CNOT}^{24}, \text{CNOT}^{34} , U^{0}(\theta) , U^{1}(\theta), U^{2}(\theta), \notag\\
& U^{3}(\theta), U^{4}(\theta )  \}\,,
\end{align}
where $U^{j}(\theta)$ is an arbitrary gate on qubit $j$ and $\text{CNOT}^{jk}$ is a $\text{CNOT}$ from control qubit $j$ to target qubit $k$. Angles $\theta$ in Eq. \eqref{eqn1} may be encoding multiple parameters. In this article, we treat all one-qubit gates equally in the sense that all one-qubit gates are equally complex to implement, although our approach could easily be generalized to account for different complexities for different one-qubit gates.

We consider a generic sequence of $d$ gates,
\begin{equation}
\label{eqn2}
G_{\vec{k}}(\vec{\theta}) =   A_{k_d}(\theta_{d}) \cdot\cdot\cdot A_{k_2}(\theta_2) A_{k_1}(\theta_1)\,,
\end{equation}
where $\vec{k} = (k_1,..., k_d)$ is the vector of indices describing which gates are employed in the gate sequence and $\vec{\theta} = (\theta_1,..., \theta_d)$ is the vector of continuous parameters associated with these gates.

The measurement results give rise to an outcome probability vector $\vec{p} = (p_1,..., p_l,...)$. The desired output might be one of these probabilities $p_l$, or it might be some simple function of these probabilities. Hence, we allow for some simple classical post-processing of $\vec{p}$ in order to reveal the desired output. While there is enormous freedom in applying a function to $\vec{p}$, we consider a simple linear combination of probabilities:
\begin{equation}
\label{eqn3}
g(\vec{p}) = \vec{c}\cdot \vec{p} = \sum_l c_l p_l
\end{equation}
where $\vec{c}$ is a vector of coefficients whose elements are chosen according to $c_l \in \{-1, 0, 1\}$. This post-processing is sufficient for the application in this paper (state overlap), although other applications may require a more general form of post-processing. Note that in our approach it is enough to consider measurements in the computational basis, as any change of the measurement basis can be incorporated into the gate sequence in Eq. \eqref{eqn2}. In particular, this implies that Eq. \eqref{eqn3} is general enough to cover the expectation values of all Pauli product operators.

In summary, the free parameters that we optimize over (while fixing the hyperparameters) are the gate sequence vector $\vec{k}$, the continuous parameter vector $\vec{\theta}$, and the post-processing vector $\vec{c}$. For a given set of resources, these three vectors define the quantum algorithm, which we denote $Q_{\vec{m}}$, where $\vec{m} = \left( \vec{k}, \vec{\theta} , \vec{c} \right)$ is the concatenated vector.

\subsection{Optimization}

Optimizing these parameters involves defining and minimizing a cost function. The cost quantifies the discrepancy between the desired output and the actual output for a given training data set.

Suppose we want to find the algorithm that computes the function $x \rightarrow f(x)$. We generate data of the form
\begin{equation}
\label{eqn4}
T = \{ (x^{(i)}, f(x^{(i)})  ) \}_{i=1}^{2N}\,.
\end{equation}
Half of this data is used for training the algorithm, i.e., optimizing the cost function. The other half is used for testing, i.e., evaluating the algorithm's performance. The training data must be sufficiently general to cover the space of possible inputs. An estimate of the amount of training data needed for state overlap is $N \approx 2^{2n_D}$, where $n_D$ is the number of data qubits. This can be seen by noting that our algorithm (which includes both the gate sequence and the post-processing) acts as a linear map from the data qubits' density operator space, which has dimension $2^{2n_D}$, to the output which is just a number and hence has dimension one. So our algorithm is basically a $1 \times 2^{2n_D}$ matrix, and an estimate of the number of constraints (and hence the number of training data points) needed to fix the algorithm's parameters is $2^{2n_D}$.

For example, when training the algorithm that computes overlap, $x^{(i)} = (\rho^{(i)}, \sigma^{(i)})$ consists of two quantum states $\rho^{(i)}$ and $\sigma^{(i)}$, and $f(x^{(i)}) = \Tr(\rho^{(i)} \sigma^{(i)} )$ quantifies their overlap. One can show that any algorithm that computes pure-state overlap also computes mixed-state overlap. Hence, we generate our training data by randomly choosing pure states according to the Haar measure.

Next we define a cost function. For algorithm $Q_{\vec{m}}$, the cost is
\begin{equation}
\label{eqn5}
C_{\vec{m}} = \sum_{i=1}^N (f(x^{(i)}) - y^{(i)}_{\vec{m}} )^2\,.
\end{equation}
The cost quantifies the difference between the ideal output $f(x^{(i)})$ and the actual output $y^{(i)}_{\vec{m}}$ for each training data point. The actual output can be written as 
\begin{equation}
\label{eqn6}
 y^{(i)}_{\vec{m}} = y^{(i)}_{(\vec{k},\vec{\theta}, \vec{c})} = \vec{c}  \cdot \vec{p}_{(\vec{k}, \vec{\theta})}^{(i)}
\end{equation}
where $\vec{c}$ is the post-processing vector and $\vec{p}_{(\vec{k}, \vec{\theta})}^{(i)}$ is the outcome probability vector for input $x^{(i)}$.
For example, in the Swap Test, the outcome probability vector corresponds to the ancilla qubit's measurement in the $Z$ basis. Choosing $\vec{c} = (1,-1 )$ ensures that $y^{(i)}_{\vec{m}}$ is the expectation value of the Pauli $Z$ operator.

\begin{figure}[t!]
\begin{center}
\includegraphics[width=\columnwidth]{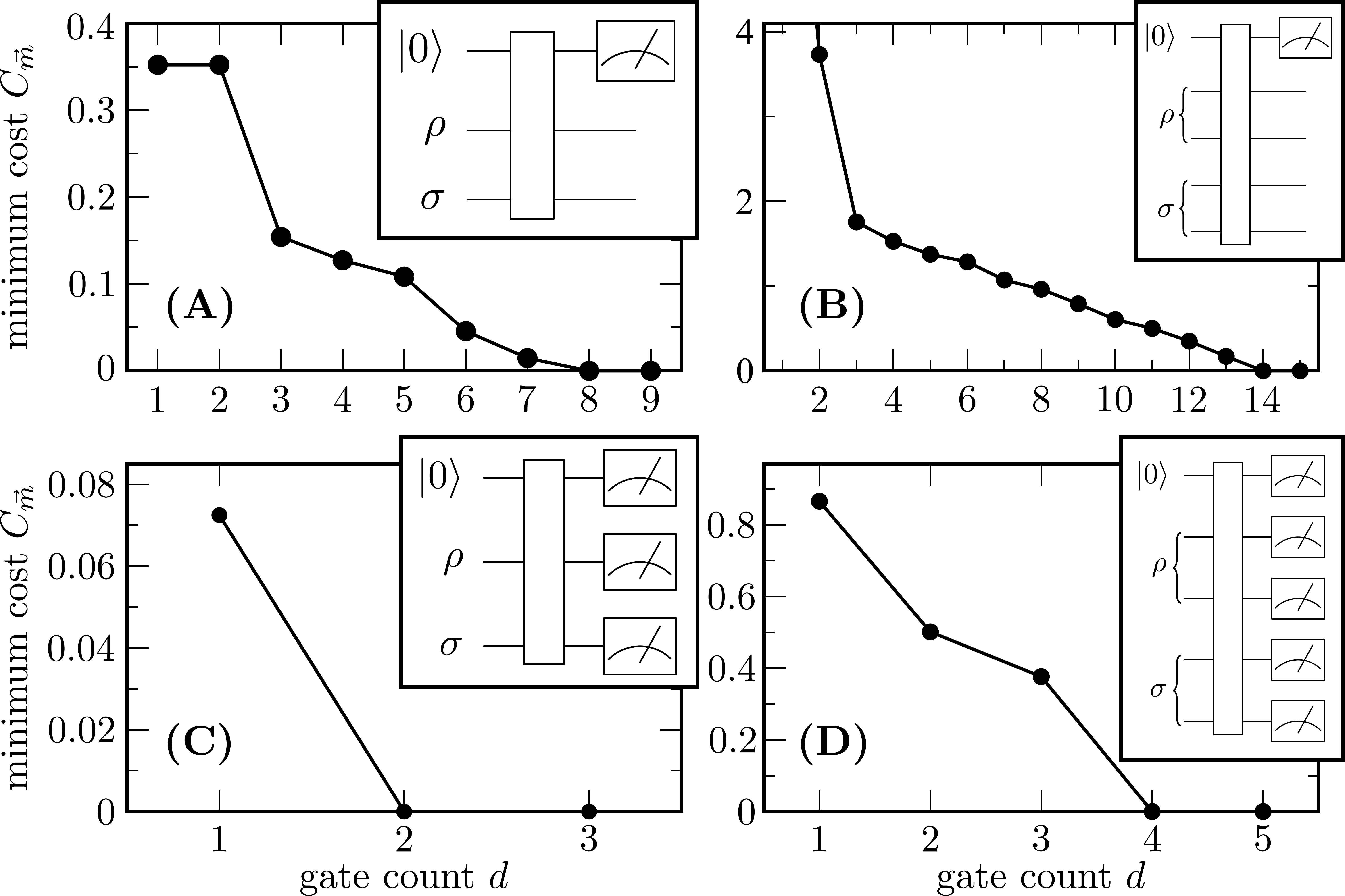}
\caption{Final cost that we obtained after minimizing our cost function versus the circuit gate count $d$. (A) The resources allowed (shown in the inset) are the same as those allowed in the Swap Test, i.e., one ancilla qubit, two data qubits, and one measurement on the ancilla. This results in a minimum gate count of $d_{\min} = 8$. (B) The number of qubits in $\rho$ and $\sigma$ is increased, resulting in $d_{\min} = 14$ for $n=2$ qubits. This procedure leads to the discovery of a general algorithm presented in Fig. \ref{figure4}. (C) Allowing for additional resources (shown in the inset) of measurements on all of the qubits results in a minimum gate count of $d_{\min} = 2$. (D) Again we increase the number of qubits in $\rho$ and $\sigma$, giving $d_{\min} = 4$ for $n=2$ qubits, when measurements on all qubits are allowed. As a result, a general algorithm is obtained, as shown in Fig. \ref{figure5}.}
\label{figure3}
\end{center}
\end{figure}

For a fixed circuit gate count $d$, we search over the algorithm space to minimize the cost, as discussed below. We consider various $d$, incrementing from small to large values. When an exact algorithm exists, we typically are able to minimize the cost. That is, we can find a $Q_{\vec{m}}$ with $C_{\vec{m}} \approx 0$, for $d\geq d_{\min}$, where $d_{\min}$ is the minimum number of gates needed to minimize the cost (see Fig.~\ref{figure3} for example plots of final cost versus $d$). Note that some elements of the gate set in Eq. \eqref{eqn1} commute with each other. As a consequence, there are typically many $Q_{\vec{m}}$ that give zero cost for $d\geq d_{\min}$. This freedom is used to simplify the algorithm at the end of the cost optimization. So, in the Main Results section, we present our simplest representation of such algorithms.

\subsection{Details of the optimization techniques}

The cost in Eq. \eqref{eqn5} is a function of several parameters that can be divided into two groups: discrete and continuous. Discrete parameters are those which describe the circuit topology and post-processing of the algorithm. These are the gate sequence vector $\vec{k}$ and the post-processing vector $\vec{c}$. The angles $\vec{\theta}$ are treated as continuous parameters. They define all gates that depend on a parameter. For IBM and Rigetti architectures considered here, angles $\vec{\theta}$ specify all one-qubit gates present in the algorithm. Only the total number of gates $d$ is fixed during optimization, which means that while the length of $\vec{k}$ does not change, the number of elements of $\vec{\theta}$ may vary as the optimization proceeds.

\begin{figure}[t!]
\begin{center}
 \includegraphics[width=\columnwidth]{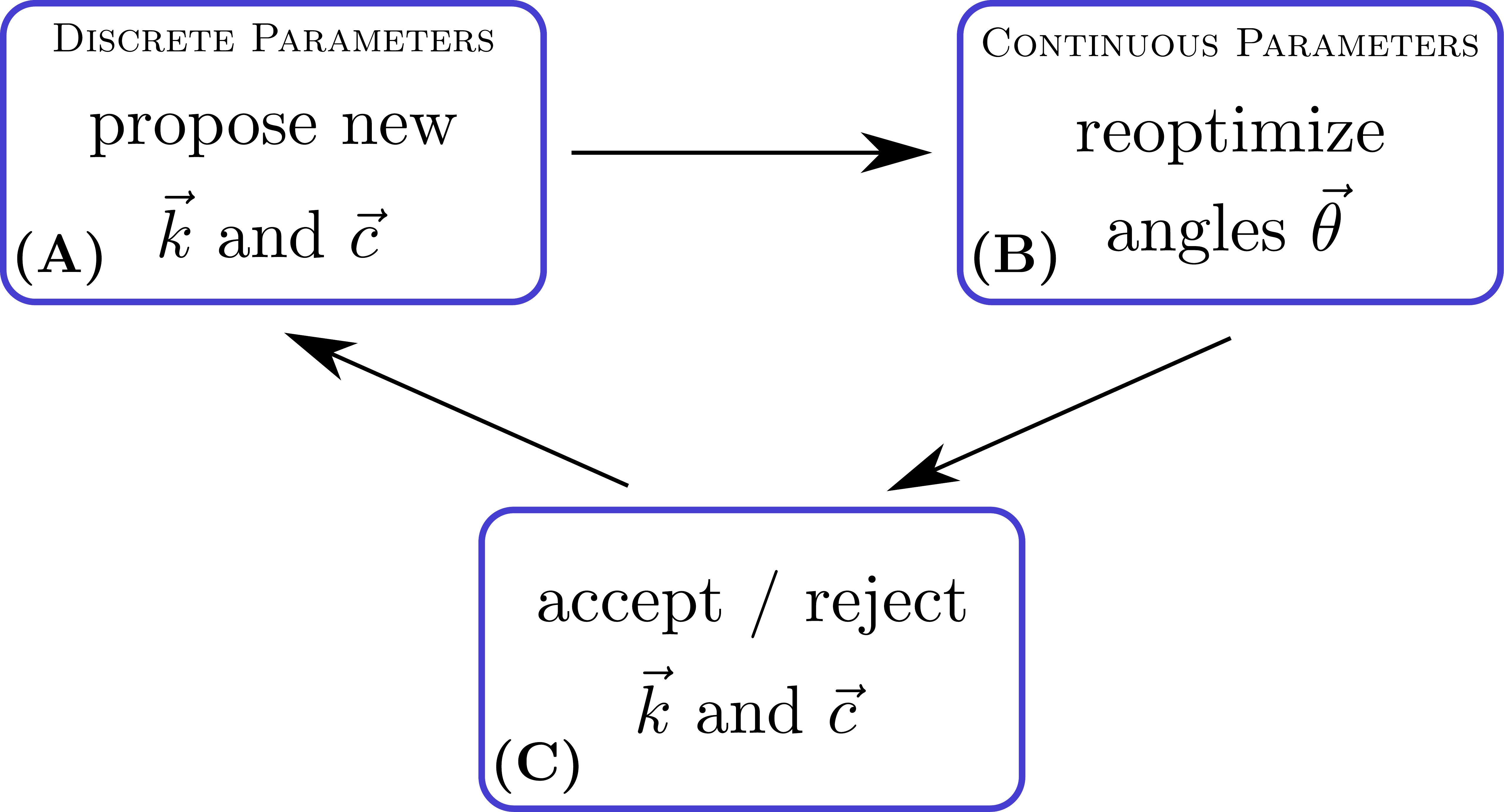}
\caption{Schematic view of one iteration of the cost optimization procedure. (A)~Iteration begins with a random update to the gate sequence vector $\vec{k}$ that describes the algorithm's structure and a random update to the post-processing vector $\vec{c}$. (B)~Continuous parameters $\vec{\theta}$ of every one-qubit gate are reoptimized using the steepest descent method. (C)~The optimization in the previous step gives a cost that is compared with the current best one. Based on the outcome of that comparison, new vectors $\vec{k}$ and $\vec{c}$ are either accepted or rejected. See text for details.}
\label{fig:details}
\end{center}
\end{figure}

The optimization is performed in iterations until the cost reaches a (possibly local) minimum. Fig. \ref{fig:details} shows a schematic description of a single iteration of the optimization algorithm. Each iteration begins with an attempt to modify $\vec{k}$ and $\vec{c}$. While modifying $\vec{k}$, we consider random updates that may involve an arbitrary number of gates. However, updates affecting a smaller number of gates are more probable. In this process, we may change the position or support of a given gate or change its type, e.g. from a one-qubit gate to a CNOT. The update is constrained to result in an algorithm that cannot be easily shortened. For example, the gate sequence that involves two one-qubit gates next to each other is not allowed, since those gates can be combined to a single one-qubit gate. This is a desired feature, as we optimize with a fixed total number of gates. Similarly, we randomly modify $\vec{c}$ giving more preference to changes affecting fewer measurements.

Every change in $\vec{k}$ or $\vec{c}$ is followed by reoptimization of continuous parameters $\vec{\theta}$. This is an important step, as changing the gate sequence or post-processing function alone, without reoptimizing the gates' internal parameters $\vec{\theta}$, will most likely cause the cost to increase significantly, effectively suppressing any update of $\vec{k}$ or $\vec{c}$. The continuous part of the optimization is done in a sweeping fashion in which all one-qubit gates are updated sequentially. In this approach, at a given time, a single one-qubit gate is updated while all remaining gates are fixed. After the best one-qubit gate (the one that minimizes the cost) is identified, the optimization algorithm moves to the next one-qubit gate. We allow for randomly changing the order of updating one-qubit gates as a means to avoid local minima. We use a steepest descent method to optimize single one-qubit gates. Note that an arbitrary one-qubit gate can be described (up to a global phase, that does not affect the algorithm) by three real parameters. That is, the steepest descent method mentioned above operates in three-dimensional space. The continuous part of the optimization is repeated, until convergence of the cost function is achieved.

Once the continuous optimization has converged, we compare the final cost $C$ in a given iteration with the current best one $C_\text{best}$. If the cost $C$ is lower than the current best, the new discrete parameters $\vec{k}$ and $\vec{c}$ are accepted. If it is larger, the change is accepted with probability exponentially decreasing in the difference $C-C_\text{best}$ following the simulated annealing method.

Every few iterations we check whether the current gate sequence $G_{\vec{k}}$ can be compressed. This goes well beyond the simple checks following the update of vector $\vec{k}$ described above. Here, we are trying to find a subsequence of $G_{\vec{k}}$ that can be nontrivially rewritten using the same or a smaller number of gates. If such a subsequence is found, we modify $G_{\vec{k}}$ accordingly, as this may lead to shortening the gate sequence without increasing the cost. Since the total number of gates is fixed, such compression results in the ability to add gates to the sequence. If that is the case, we insert one-qubit identity gates and reoptimize their continuous parameters as described above. To check if a given subsequence can be rewritten we recursively use the same approach that we use for the full algorithm, which is essentially described in Fig. \ref{fig:details} except in this case we do not consider the post-processing vector $\vec{c}$.

We remark that the cost function may be difficult to optimize primarily due to many low lying local minima. Thus, it is important to develop techniques to increase the chances of avoiding them. We found it particularly useful to compress the gate sequence periodically, as random updates to vectors $\vec{k}$ and $\vec{c}$ tend to produce local minima that usually include redundant subsequences. As described above, we have developed automated tools to remove such subsequences, which usually allows us to escape local minima.

Let us now discuss the scaling of the approach described above. The optimization requires the cost to be evaluated multiple times during every iteration. As part of computing the cost, one has to evaluate $y^{(i)}_{\vec{m}}$ in Eq. \eqref{eqn5} for each training data point, which necessarily scales exponentially with the number of qubits on a classical computer. However, it can be outsourced to a quantum computer. Such a hybrid algorithm will efficiently compute the contribution to the cost from a single element of a training data set, although the resulting cost will reflect the quantum hardware's noise. In this work, we evaluate the cost on a classical computer, as we are mainly interested in the discovery of theoretical algorithms without device-specific noise considerations.

Another aspect of the algorithm scaling is training data. In general, its size will scale exponentially with the number of data qubits. However, we would like to stress that this fact does not jeopardize our approach since we numerically obtain solutions (algorithm instances) only for a small number of data qubits. Those optimization problems require training data that is still manageable. Algorithm instances are then used to manually recognize the pattern and generalize to arbitrary system size.

Finally, the search space defined by $\vec{k}$ is exponentially large in the number of gates. This makes it impossible to systematically check all possibilities in the search for an optimal algorithm. On the other hand, the heuristic approach described above seems to be capable of finding the solution efficiently.  

\subsection{Generalization}

For a fixed problem size, we minimize the cost. If the cost goes to zero (which we define as a cost less than $10^{-6}$), we say we have an {\it algorithm instance}. In particular, this corresponds to fixing the size of the data and hence fixing $n_D$, the number of data qubits. To study the generalization of the algorithm, we grow the size of the problem by increasing $n_D$. In some cases, one may also need to increase the number of ancilla qubits, $n_A$, and/or the number of measurements in order to minimize the cost.

This gives us a set of algorithm instances for various problem sizes. An important challenge is to abstract a general algorithm from these instances. This challenge is particularly difficult because one can typically only find algorithm instances for small problem sizes. This is due to the fact that the search space for vectors $\vec{k}$ grows rapidly with problem size, namely as $n_T^{2d}$, where $n_T = n_D + n_A$ is the total number of qubits and $d$ is the circuit gate count.

In this work, we were able to manually recognize the pattern by which the algorithm generalizes to arbitrary problem size by inspecting the various algorithm instances. In future work, we will explore automated methods for recognizing the general algorithm.

\section{Main Results}\label{sctmainresult}

\subsection{Overview}

Our main results are short-depth algorithms for quantifying overlap on idealized quantum computing hardware. For the latter, we consider full connectivity, and we allow for arbitrary one-qubit gates as well as $\text{CNOT}$ gates between all of the qubits.

\begin{figure}
\begin{center}
 \includegraphics[width=\columnwidth]{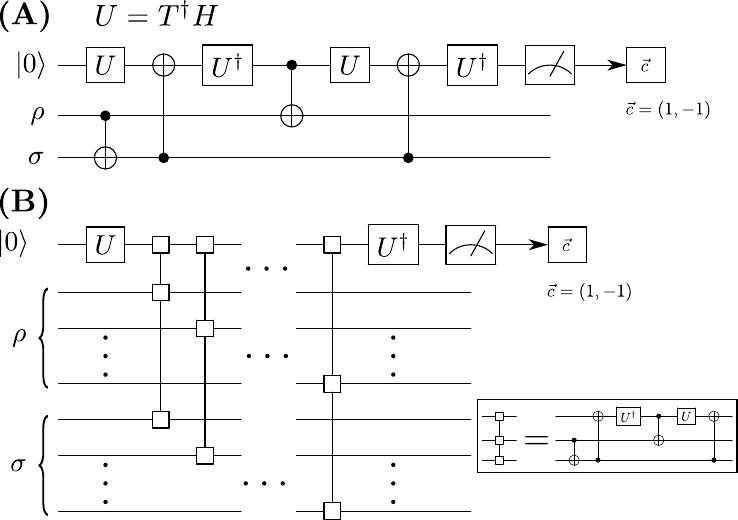}
\caption{Our Ancilla-Based Algorithm (ABA), obtained by minimizing the cost for the resources shown in Fig.~\ref{figure3}(A) and (B). (A) When $\rho$ and $\sigma$ are one-qubit states, we obtain a circuit with 4 $\text{CNOT}$ gates and 4 one-qubit gates for a total of 8 gates. Here, $U =  T\ad H$. (B) Six of these gates are combined to create a ``building block'' (see inset) that is used to generalize the algorithm for input states $\rho$ and $\sigma$ of arbitrary size. The post-processing vector is $\vec{c} = (1, -1)$, independent of problem size.}
\label{figure4}
\end{center}
\end{figure}

We consider two sets of resources. The first set of resources are identical to those allowed for the Swap Test, i.e., access to one ancilla qubit and two data qubits, as well as one measurement on the ancilla qubit. The cost versus number of gates for these resources is shown in Fig.~\ref{figure3}(A), and we obtained essentially zero cost for $d=8$. To understand how the algorithm generalizes, we increase the number of qubits in $\rho$ and $\sigma$ to $n=2$, giving a minimum gate count of $d=14$, as shown in Fig.~\ref{figure3}(B). As discussed below this generalizes to an algorithm (shown in Fig.~\ref{figure4}) that we refer to as our Ancilla-Based Algorithm (ABA). 

\begin{figure}[t!]
\begin{center}
 \includegraphics[width=0.9\columnwidth]{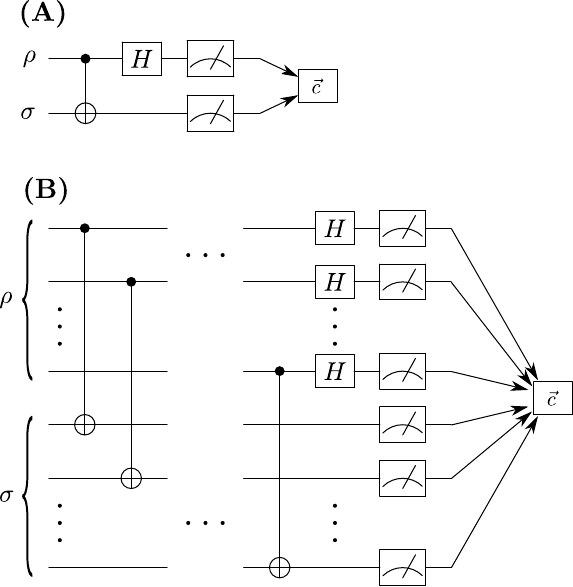}
\caption{Our Bell-Basis Algorithm, obtained by minimizing the cost for the resources shown in Fig.~\ref{figure3}(C) and (D). (A) When $\rho$ and $\sigma$ are one-qubit states, we obtain a circuit with one $\text{CNOT}$ followed by a Hadamard and measurements on both qubits with a post-processing vector $\vec{c} = (1, 1, 1, -1)$. (B) The $\text{CNOT}$ and Hadamard gates form a ``building block'' that is used to generalize the algorithm for input states $\rho$ and $\sigma$ of arbitrary size. Since these gates can be parallelized, the quantum circuit depth is independent of problem size. On the other hand, the complexity of classical post-processing grows linearly with $n$, and the post-processing vector can be written as $\vec{c} = (1, 1, 1, -1)^{\otimes n }$ if one orders the qubits into pairs from $\rho$ and $\sigma$.}
\label{figure5}
\end{center}
\end{figure}

The second set of resources we consider allows for measurements on all of the qubits. For these additional resources, Figure~\ref{figure3}(C) shows that zero cost is obtained for $d = 2$. To recognize how this algorithm generalizes, we increase the number of qubits to $n=2$, giving a minimum gate count of $d=4$, as shown in Fig.~\ref{figure3}(D). The surprising result is that the ancilla qubit is not used at all in this algorithm, even though we train the algorithm in the presence of an ancilla. This allows us to display the resulting general algorithm, our Bell-Basis Algorithm (BBA), in Fig.~\ref{figure5} without the ancilla qubit. 

In both cases discussed above, we managed to discover the general (valid for arbitrary problem size) form of the algorithm from its two smallest instances. We expect that in other applications, the general form of the algorithm may be harder to find and more sophisticated tools will have to be developed.

\subsection{Ancilla-Based Algorithm}

Figure~\ref{figure4}(A) shows the ABA for one-qubit states $\rho$ and $\sigma$. The unitary $U$ in this circuit is $U =  T\ad H$. This circuit employs 4 $\text{CNOT}$ gates and 4 one-qubit gates for a total of 8 gates. It uses a simple post-processing vector $\vec{c} = (1,-1)$ that amounts to measuring the Pauli $Z$ operator on the ancilla qubit, which is the same observable measured in the Swap Test. Not only does this circuit have a lower gate count than typical implementations of the Swap Test (see e.g., the circuit in Fig.~\ref{figure1}(B)), but actually it implements a completely different unitary.

Let $S_{\text{ABA}}$ denote the Schmidt rank (across the cut between ancilla and the data qubits) of the unitary $G_{\text{ABA}}$ associated with the ABA gate sequence. It can be verified that $S_{\text{ABA}} = 3$. This means that $G_{\text{ABA}}$ is not locally equivalent to a controlled-SWAP, whose analogously defined Schmidt rank is 2. Thus, the ABA is fundamentally different from the Swap Test: it cannot be obtained from the Swap Test by local operations.

The general form of the ABA is given in Fig.~\ref{figure4}(B). There is a repeating unit, shown in the inset of the figure, that is applied on each pair of qubits composing $\rho$ and $\sigma$ as well as on the ancilla qubit. This unit has 4 $\text{CNOT}$ gates, so the overall algorithm employs $4n$ $\text{CNOT}$ gates and $6n +2$ total gates. Hence, the gate count grows linearly with the number of data qubits.

\subsection{Bell-Basis Algorithm}

Figure~\ref{figure5}(A) shows the BBA for one-qubit states $\rho$ and $\sigma$. This circuit employs one $\text{CNOT}$ gate followed by one Hadamard gate, with both qubits being measured. It is straightforward to show that this corresponds to a Bell basis measurement. The post-processing is a bit more complicated, with $\vec{c} = (1, 1, 1, -1)$, which corresponds to summing the probabilities for the 00, 01, and 10 outcomes and subtracting probability of the 11 outcome. The above  post-processing is equivalent to measuring the expectation value of a controlled-$Z$ operator.

\begin{figure}[t!]
\begin{center}
 \includegraphics[width=1.00\columnwidth]{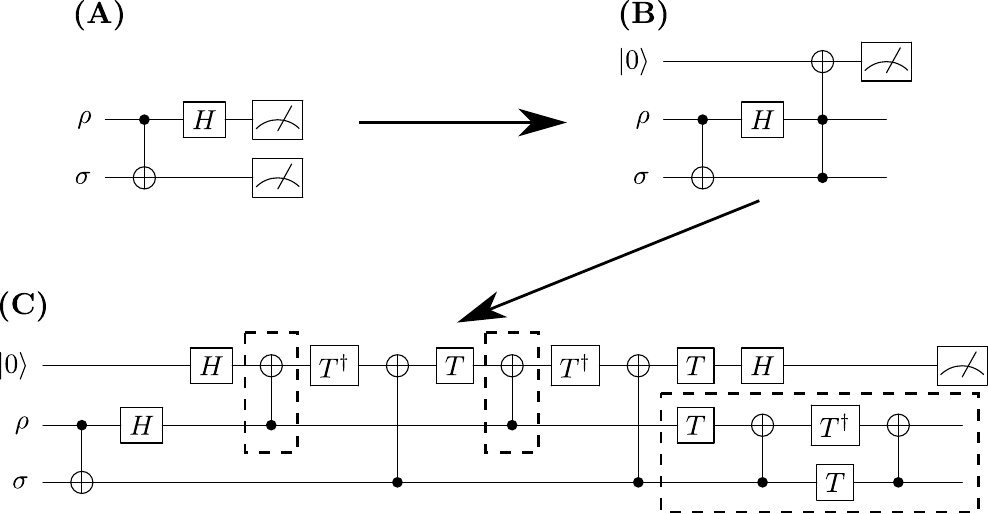}
\caption{Equivalence between our ABA and BBA. The two-qubit measurement and classical post-processing in the BBA can be converted to a Toffoli gate with an ancilla as the target followed by a measurement on the ancilla. This takes us from circuit (A) to circuit (B). Inserting into circuit (B) the optimal decomposition of the Toffoli gate from Ref.~\cite{shende2009cnot} gives circuit (C). Finally one does three simplifications of this circuit to obtain the ABA, indicated by the dashed boxes in (C). Namely, the first boxed CNOT in (C) has trivial action and hence can be removed. The second boxed CNOT in (C) can be flipped such that the ancilla is the control qubit, which introduces some Hadamards. One of these Hadamards cancels with the first Hadamard in (C), and two others combine with $T$ and $T\ad$ to make the $U\ad$ and $U$ shown in Fig.~\ref{figure4}(A). Finally the five gates enclosed in the last dashed box in (C) have no effect on the measurement and hence can be removed.}
\label{figure6}
\end{center}
\end{figure}

The generalization of this algorithm is given in Fig.~\ref{figure5}(B). The repeating unit is simply a $\text{CNOT}$ and Hadamard, applied on each pair of qubits composing $\rho$ and $\sigma$. Furthermore, every qubit is measured at the output. The total number of gates is simply $2n$, and hence grows linearly with the number of qubits. However, more importantly, the $\text{CNOT}$ and Hadamard on each qubit pair can be performed in parallel. This crucial fact means that this algorithm has a constant depth, independent of problem size. Namely, the depth is two quantum gates.

On the other hand, the classical post-processing is somewhat complicated, and its complexity scales linearly with the problem size. Namely, the post-processing vector can be written as $\vec{c} = (1, 1, 1, -1)^{\otimes n  }$, provided that one arranges the qubits in the order $P_1 Q_1 P_2 Q_2 ....P_n Q_n$, where $P_1 P_2 ...P_n$ and $Q_1 Q_2... Q_n$ are the subsystems composing $\rho$ and $\sigma$ respectively. The linear scaling of post-processing follows from the fact that one does not explicitly compute $\vec{c} \cdot \vec{p}$ in Eq. \eqref{eqn3}. Rather one bins individual measurement outcomes into one of two bins (either the $1$ or $-1$ bin). Here, the bin is determined by first assigning each of the $n$ qubit pairs a value of $1$ or $-1$, based on the associated eigenvalue of the controlled-$Z$ operator, and then multiplying these $n$ values. The overlap $\Tr(\rho\sigma)$ is then given by the weighted average over all outcomes, where the weights correspond to the bin label (either $1$ or $-1$).

Nevertheless, for NTQCs, due to decoherence and gate infidelity, it is better for the classical post-processing to grow linearly in $n$ than for the quantum circuit depth to grow linearly in $n$. Hence, the BBA seems to be the superior algorithm in that case.

\subsection{Discussion} \label{sec:discussion}

In 2013, Garcia-Escartin and Chamorro-Posada discovered the Bell-Basis Algorithm for computing state overlap \cite{garcia2013swap}. We were unaware of this important result until after our machine-learning approach found our BBA. More generally, it appears that the quantum computing community seems to be unaware of this article, perhaps because the article was presented in the language of quantum optics rather than that of quantum computing. Indeed, the ancilla-based version of the Swap Test, shown in Fig.~\ref{figure1}, continues to be the algorithm employed in the quantum computing literature (e.g., see Refs.~\cite{linke2017measuring, johri2017entanglement}).

Although our two algorithms look very different, one can actually show a simple equivalence between our ABA and our BBA. One can see this by converting the classical post-processing in the BBA into a quantum gate. In particular, this gate would be a Toffoli gate, controlled by the two data qubits with the target being an ancilla qubit prepared in the $\ket{0}$ state. Appendix \ref{sctequiv} shows proof of this statement. After inserting the Toffoli gate (see Fig. \ref{figure6}(B)), one would do a measurement of the Pauli $Z$ observable on the ancilla to decode the state overlap. By replacing the Toffoli gate with its decomposition from Ref.~\cite{shende2009cnot} and simplifying the resulting circuit, one can then obtain our ABA (see Fig. \ref{figure6}(C)). In this sense, our ABA is essentially our BBA but with the classical post-processing transformed into Toffoli gates and a measurement on the ancilla. This equivalence is shown in Fig.~\ref{figure6} for one-qubit states. The generalization to multi-qubit states is straightforward.

\begin{figure}[t!]
\begin{center}
 \includegraphics[width=1.00\columnwidth]{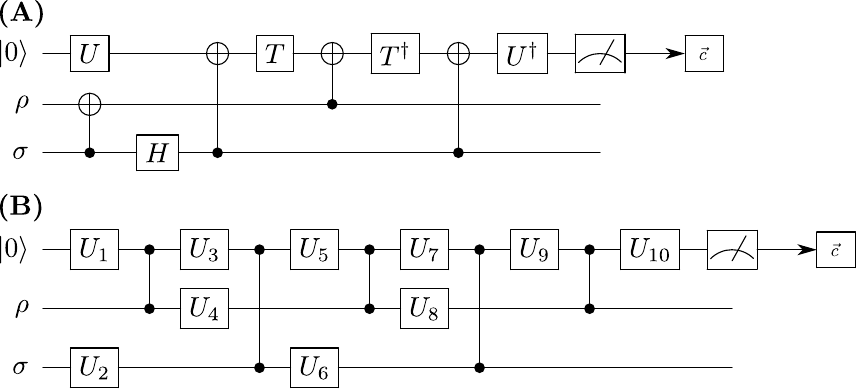}
\caption{Ancilla-based algorithm adapted (via our machine-learning approach) to commercial hardware. (A) ABA adapted to IBM's 5-qubit computer, $U=T^\dagger H$. (B) ABA adapted to Rigetti's 19-qubit computer. One-qubit unitaries have the following form: $U_1 = U_8 = H$, $U_2 = U_3 = U_6^\dagger = U_7^\dagger = XH$, $U_4 = R_X(-\pi/4)T$, $U_5 = T^\dagger HT$, $U_9 = R_X(\pi/4)$, $U_{10} = R_X(-3\pi/4)$, where $R_X(\theta) = e^{-i\frac{\theta}{2} X}$.
}
\label{figure7}
\end{center}
\end{figure}

\section{Hardware-specific algorithms}\label{sctbackground}

Our BBA can be directly implemented on IBM's and Rigetti's quantum computers without any concern about connectivity issues (except for the minor issue that Rigetti uses controlled-$Z$ instead of CNOT - their compiler easily makes the translation). 

However, our ABA needs to be modified to account for IBM's and Rigetti's connectivity. While it is possible to manually modify the ABA to fit the connectivity, to illustrate our machine-learning approach, we numerically optimized the algorithm with the same resources as that shown in Fig.~\ref{figure3}(A). The only difference is that we specified the gate set $\AC$ to match the gate set (and hence the connectivity) of IBM's and Rigetti's computers. 

The resulting algorithms that we obtained with our machine-learning approach are shown in Fig.~\ref{figure7}. The ABA adapted to IBM's 5-qubit computer only requires one additional gate, a Hadamard gate. The ABA adapted to Rigetti's 19-qubit computer requires an additional two-qubit gate and several additional one-qubit gates.

\begin{figure}[t!]
\begin{center}
 \includegraphics[width=1.0\columnwidth]{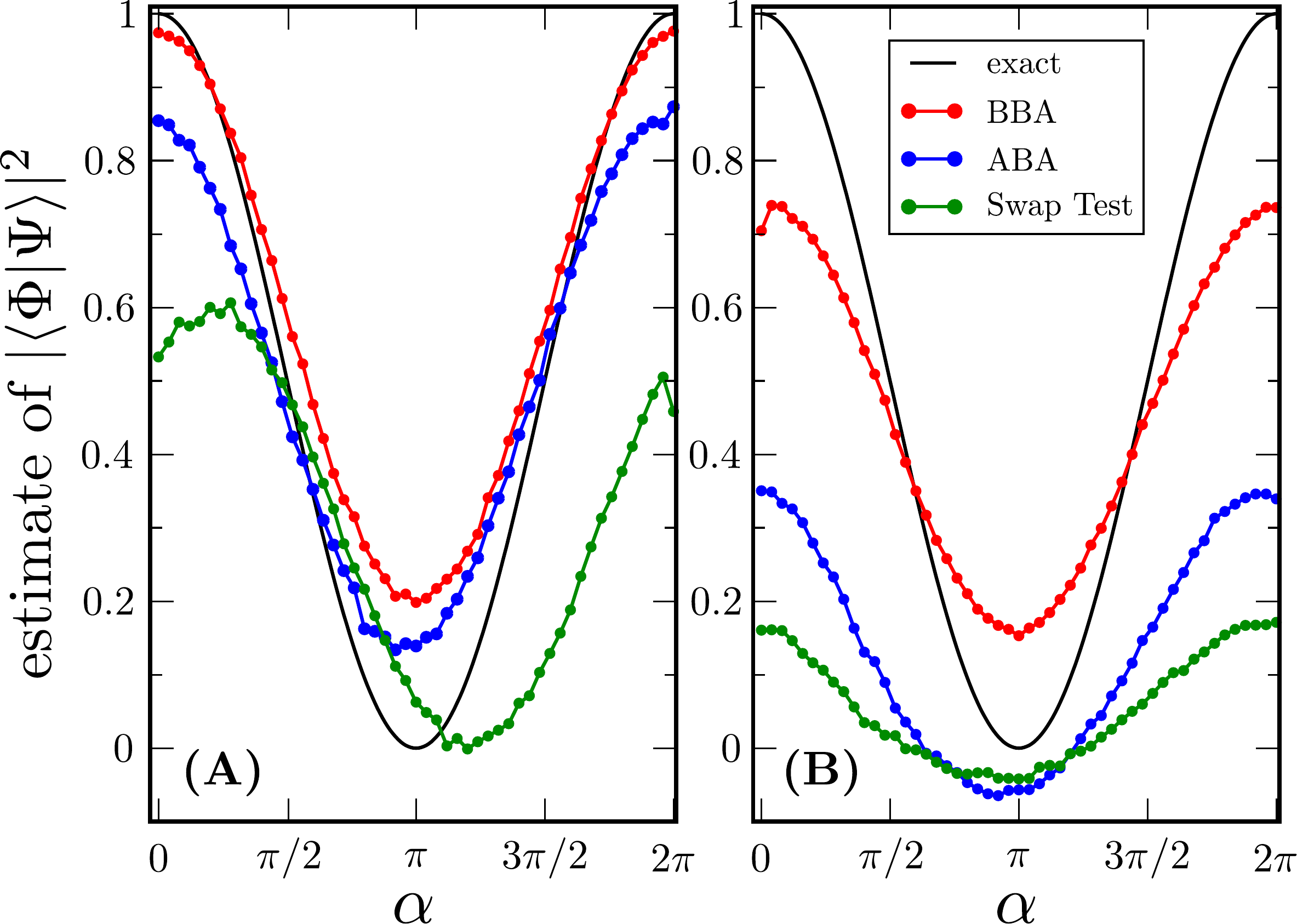}
\caption{Experimentally observed overlaps on commercial hardware for the states $\ket{\Psi} = (\ket{0}+ \ket{1})/\sqrt{2}$ and $\ket{\Phi} = (\ket{0}+ e^{i\alpha}\ket{1})/\sqrt{2}$.  (A) Results from IBM's 5-qubit computer called ``ibmqx4'', with 49,152 quantum computer runs per data point. The black curve is the analytical overlap $|\ip{\Phi}{\Psi}|^2$. The red, blue, and green curves are respectively the results for the BBA from Fig.~\ref{figure5}(A), the ABA from Fig.~\ref{figure7}(A), and the Swap Test from Fig.~\ref{figure2}(B).  (B) Results from Rigetti's 19-qubit computer, with 200,000 quantum computer runs per data point. The curves are analogous to those from panel (A). Namely, the red, blue, and green curves are respectively for the BBA from Fig.~\ref{figure5}(A), the ABA from Fig.~\ref{figure7}(B), and the Swap Test from Fig.~\ref{figure2}(A) which Rigetti compiled to Fig.~\ref{figure2}(C). The experimentally estimated overlap takes negative values for some $\alpha$ because the algorithm estimates the expectation value of controlled-$Z$ operator, which has a negative eigenvalue. Another reason for this effect may be noise and other imperfections of the device.
}
\label{figure8}
\end{center}
\end{figure}

\section{Testing our algorithms}\label{sctbackground2}

We implemented our algorithms on IBM's 5-qubit and Rigetti's 19-qubit computers. The resulting data are shown in Fig.~\ref{figure8}. A caveat is that the different qubit counts for the two hardwares make it difficult to directly compare the results between these hardwares.

We considered two pure states of the form
\begin{align}
\label{eqn8}
\ket{\Psi} &= (\ket{0}+ \ket{1})/\sqrt{2}\\
\ket{\Phi} &= (\ket{0}+ e^{i\alpha}\ket{1})/\sqrt{2}\,, \label{eq:Phi}
\end{align}
and we compared our results to the exact overlap $|\ip{\Phi}{\Psi}|^2$ (black curve in Fig.~\ref{figure8}). The root-mean-square (RMS) errors are shown in Table~\ref{table1}.

On both computers, the Swap Test (green curve in Fig.~\ref{figure8}) performed poorly. It is noteworthy that these are only single-qubit states, and hence the results are expected to be even worse as one grows the size of these states.

Overall, our ABA performed significantly better than the Swap Test, while using the same resources, as is evident from the much smaller RMS errors. The BBA, which allows for measurements on all qubits, dramatically outperformed the other algorithms on Rigetti's computer and performed roughly the same as ABA on IBM's computer. The relatively high accuracy of BBA is naturally expected due to its short depth, which mitigates the effects of decoherence and gate infidelity. 

We note that there are values of the parameter $\alpha$ in Eq. \eqref{eq:Phi} for which the Swap Test performs better than ABA and BBA, e.g. around $\alpha \approx \pi$. However, we believe that the RMS error given in Table \ref{table1} is a better indicator of algorithms' performance than the error at a particular value of $\alpha$. To make this point, note that on a fully decohered (but otherwise perfect) hardware, the Swap Test is expected to return zero overlap independently of angle $\alpha$. The algorithm would output the correct value for the overlap at $\alpha = \pi$ albeit for the wrong reason.

\begin{table}[t!]
\centering
\begin{tabular}{ |c|c|c| } 
 \hline
  & IBM (5 qubits) & Rigetti (19 qubits) \\ 
 \hline
 Swap Test & 0.311   & 0.537 \\
 ABA           & 0.106  & 0.432 \\
 BBA           & 0.116  & 0.160 \\
 \hline
\end{tabular}
\caption{RMS errors for the data shown in Fig.~\ref{figure8}.}
\label{table1}
\end{table}

Our results show that connectivity between qubits as well as native gate set play important roles in the performance. Rigetti's 19 qubit computer offers less connectivity than IBM's 5-qubit one. As a result, algorithms discovered for Rigetti's architecture are longer (compare circuits presented in Figs. \ref{figure2} and \ref{figure7}) and overall perform worse. Algorithms found for IBM and Rigetti's computers suggest that for the particular problem of finding $\mathrm{Tr}(\rho \sigma)$, the ability to apply CNOT (rather that controlled-Z) results in shorter circuits. This can be seen from Fig. \ref{figure7}(B): several one-qubit gates can be eliminated by writing controlled-Z gates in terms of CNOTs.

\section{Conclusions}\label{sctconclusion}

This work shows that even well-known algorithms can be improved upon using an automated approach.  As noted in the Introduction, there are many applications that require state overlap computation, including the emerging new field of quantum machine learning. While the Swap Test appears as a subroutine in many of these applications, we show that there are more efficient circuits to perform this subroutine.

We have found a constant depth algorithm (denoted BBA above) for computing state overlap, which is better than the linear scaling of the Swap Test. Furthermore, this algorithm performs better - with significantly lower error - even in the single-qubit case. It is therefore advisable that researchers use this algorithm henceforth for computing state overlap on NTQCs. This algorithm essentially corresponds to a measurement in the Bell basis for corresponding pairs of qubits. A key aspect of our approach that aided this algorithm's discovery was to allow for non-trivial classical post-processing, a strategy that has been used previously to shrink the depth of quantum algorithms \cite{svore2014faster}. The complexity of the post-processing for the BBA scales only linearly in the problem size (i.e., the number of qubits), ensuring that the quantum speedup that this algorithm provides is not due to the transfer of exponential complexity to the classical post-processing, but rather comes from the use of gates that can be executed in parallel.

Our main technical tool was a machine-learning method that allowed for task-oriented discovery of quantum algorithms. By task-oriented, we mean that this method defines a cost function based upon training data that are representative of the desired computation, i.e., the training data define the task. Minimizing the cost function results in a general algorithm for this computation. We emphasize that this goes far beyond quantum compiling since it allows for algorithm discovery when no algorithm is known.

Conceptually, our method separates quantum resources (ancillas, data qubits, and measurements) from algorithm parameters (gate sequence and classical post-processing). The former are fixed as hyperparameters while we optimize the latter. The algorithm's generalization is obtained by training for various problem sizes and recognizing the pattern. In future work, we plan to automate the process of pattern recognition for algorithm generalization.

As noted in \cite{gepp2009review}, this field will be even more promising when quantum computers become available. This is due to the exponential speedup they provide in evaluating algorithm cost, i.e., by avoiding the exponential overhead of quantum simulation on classical computers. Indeed, some recent works propose to use quantum computers in automated algorithm learning \cite{mitarai2018quantum, khatri2018quantum, benedetti2018generative}. Likewise our method can be extended to learning on a quantum computer by outsourcing cost evaluation to the quantum computer. This will be a topic of our future work. 

\section{Acknowledgements}\label{sctackknowledge}

The authors acknowledge helpful discussions with Francesco Caravelli. We thank Rigetti and IBM for providing access to their quantum computers. The views expressed in this paper are those of the authors and do not reflect those of Rigetti or IBM. LC was supported by the U.S. Department of Energy through the J.~Robert Oppenheimer fellowship. YS acknowledges support of the LDRD program at Los Alamos National Laboratory (LANL). ATS and PJC were supported by the LANL ASC Beyond Moore's Law project. 

\appendix

\section{Implementation details} \label{sctdetails}

This Appendix gives details on the implementation of the Swap Test on Rigetti's 19-qubit quantum computer. The circuit, shown in Fig. \ref{fig:Rigetti_details}, was generated by Rigetti's compiler. It consists of 22 one-qubit gates decomposed into rotations $R_Z(\alpha) = e^{-i \frac{\alpha}{2} Z}$ and pulses $S = e^{-i\frac{\pi}{4} X}$ as follows:

\begin{figure}
\begin{center}
 \includegraphics[width=1.0\columnwidth]{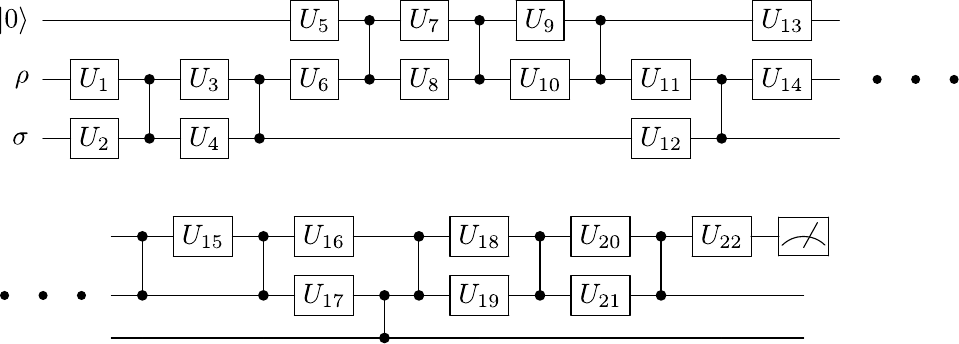}
\caption{Swap Test circuit obtained from Rigetti's compiler for their 19-qubit quantum computer. The specific form of all one-qubit gates is given by Eq. \eqref{eq:1q_gates}.  }
\label{fig:Rigetti_details}
\end{center}
\end{figure}

\begingroup
\allowdisplaybreaks
\begin{align*} 
        U_1 & = U_2 = S R_Z(-3\pi/4) S \ , \numberthis \label{eq:1q_gates} \\
        U_3 & = S R_Z(-\pi/2) \ ,   \\
        U_4 & = S^\dagger R_Z(\pi/4) S R_Z(\pi/2) \ , \\
        U_5 & = S R_Z(\alpha_1) S R_Z(-\pi/2) \ ,  \\
        U_6 & = S^\dagger R_Z(\alpha_2) S R_Z(3\pi/4) \ ,  \\
        U_7 & = S^\dagger R_Z(-\pi/2) \ ,  \\
        U_8 & = U_9 = U_{12} = U_{14}^\dagger = U_{18}^\dagger = U_{21}^\dagger =  S \ ,   \\
        U_{10} & = S^\dagger R_Z(\pi/4) S R_Z(-\pi/2) \ ,  \\
        U_{11} & = S^\dagger R_Z(\alpha_3) S \ ,  \\
        U_{13} & = S^\dagger R_Z(\alpha_4) \ ,   \\
        U_{15} & = S R_Z(\pi/4) S^\dagger R_Z(\pi) \ ,  \\
        U_{16} & = S R_Z(-3\pi/4) S R_Z(\pi/2) \ ,  \\
        U_{17} & = S R_Z(-\pi/4) \ ,  \\
        U_{19} & = U_{20} = S R_Z(\pi) \ ,   \\
        U_{22} & = R_Z(-\pi/2) S R_Z(\pi/4) \ ,  \\
\end{align*} 
\endgroup
where $\alpha_1 \simeq -0.6544\pi$, $\alpha_2 \simeq 0.7857\pi$, $\alpha_3 \simeq 0.1544\pi$ and $\alpha_4 \simeq 0.2143\pi$.

\section{Equivalence between ABA and BBA} \label{sctequiv}

\begin{figure}
\begin{center}
 \includegraphics[width=1.0\columnwidth]{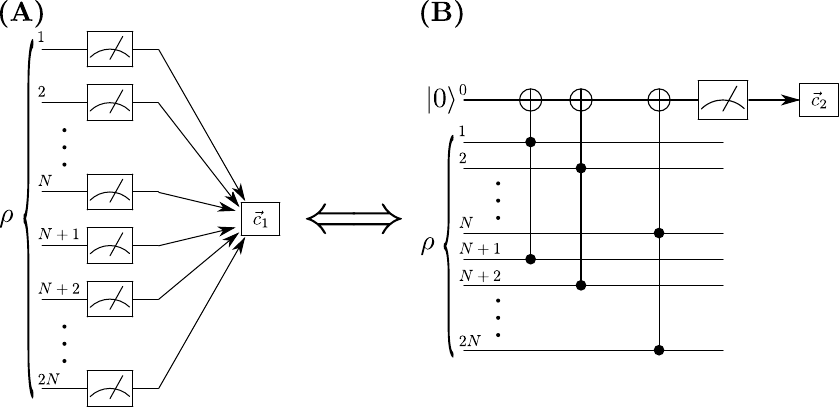}
\caption{Post-processing that is used in BBA (panel (A)) is equivalent to the sequence of Toffoli gates followed by a measurement of Pauli $Z$ operator on ancilla qubit shown in panel (B). Here the post-processing vectors are $\vec{c}_2 = (1,-1)$ and $\vec{c}_1 = (1,1,1,-1)^{\otimes n}$ , assuming qubits are arranged in order $1,N+1,2,N+2, \ldots ,N,2N$. }
\label{fig:equiv}
\end{center}
\end{figure}

Here we show that the post-processing in the BBA is equivalent to inserting a sequence of Toffoli gates followed by a measurement of Pauli $Z$ operator as shown in Fig. \ref{fig:equiv}. The rest of the proof of equivalence between ABA and BBA is presented in section \ref{sec:discussion} for one-qubit input states. Generalization to multi-qubit input states is straightforward as Toffoli gates in Fig. \ref{fig:equiv} are controlled by different qubits.

Let $\mathrm{CZ}_{j,k}$ denote controlled-$Z$ gate acting on qubits $j$ and $k$. Note that CZ is symmetric - the roles of control and target qubits can be exchanged. Post-processing employed in BBA is equivalent to measuring the expectation value of a product of CZ gates. The outcome of BBA is thus given by
\begin{equation}
    \Tr \left[ \rho \prod_{k=1}^N \CZ_{N+k,k} \right] \ ,
\end{equation}
where $\rho$ is $2N$-qubit density matrix describing the state of BBA just before the measurement, see Fig. \ref{fig:equiv}(A). We will show that this quantity is equal to the outcome of the algorithm that is obtained from BBA by replacing measurement on all qubits and subsequent post-processing with a collection of Toffoli gates followed by measurement on the ancilla qubit, as shown in Fig. \ref{fig:equiv}(B). The outcome of that algorithm is given by
\begin{equation} \label{eq:toffoli}
    \Tr \left[
    \prod_{k=1}^N T_{N+k,k,0}
    ( \ketbra{0}{0} \otimes \rho ) 
    \prod_{k=1}^N T_{N+k,k,0} 
    Z_0
    \right] ,
\end{equation}
where $T_{j,k,0}$ denotes Toffoli gate acting on qubits $j,k,0$ with $j,k$ being control qubits and $0$ is the target qubit. $Z_0$ denotes Pauli $Z$ operator acting on qubit $0$. The expression in \eqref{eq:toffoli} can be transformed as follows

\begingroup
\allowdisplaybreaks
\begin{align*}
& \Tr \left[  ( \ketbra{0}{0} \otimes \rho )
\prod_{k=1}^N T_{N+k,k,0} 
Z_0
\prod_{k=1}^N T_{N+k,k,0} \right] = \\
& \Tr \left[  ( \ketbra{0}{0} \otimes \rho )
\CZ_{1,N+1}
\prod_{k=2}^N T_{N+k,k,0} 
Z_0 
\prod_{k=2}^N T_{N+k,k,0} \right] = \\
& \ldots \\
& \Tr \left[  ( \ketbra{0}{0} \otimes \rho )
Z_0 \prod_{k=1}^N \CZ_{N+k,k} \right] = \\
& \Tr \left[ \rho \prod_{k=1}^N \CZ_{N+k,k} \right] \ , \numberthis \label{eq:deriv}
\end{align*}
where we used the fact that $T_{k,j,0}$ commutes with $T_{k',j',0}$, as well as $\CZ_{k',j'}$. We also used the following gate equivalence
\begin{equation}
    T_{k,j,0} Z_0 T_{k,j,0} = Z_0 \CZ_{k,j} \ .
\end{equation}

The last line in Eq. \eqref{eq:deriv} establishes the equivalence.

\bibliographystyle{naturemag}
\bibliography{overlap}

\end{document}